\documentclass[%
reprint,
amsmath,amssymb,
aps,
]{revtex4-1}

\usepackage{graphicx}
\usepackage{float}
\usepackage{wrapfig}
\usepackage{dcolumn}
\usepackage{bm}

\usepackage{lineno,hyperref}
\usepackage{epsfig}
\usepackage{float}
\usepackage{bookmark}
\usepackage{times}
\usepackage{amsmath}
\usepackage{color}
\usepackage{epstopdf}
\allowdisplaybreaks[4]
\usepackage{stfloats}

\begin{document}
	
	\preprint{APS/123-QED}
	
	\title{Evolutionary dynamics of continuous public goods games in structured populations}
	
	\author{Jing Luo}
	
	\author{Duozi Lin}
	
	\author{Xiaojie Chen}
	\email{xiaojiechen@uestc.edu.cn (Corresponding author)}
	\affiliation{School of Mathematical Sciences, University of Electronic Science and Technology of China, Chengdu 611731, China}
	
	\author{Attila Szolnoki}
	\affiliation{Institute of Technical Physics and Materials Science, Centre for Energy Research, P.O. Box 49, H-1525 Budapest, Hungary}

	\begin{abstract}
		Over the past few decades, many works have studied the evolutionary dynamics of continuous games. However, previous works have primarily focused on two-player games with pairwise interactions. Indeed, group interactions rather than pairwise interactions are usually found in real situations. The public goods game serves as a paradigm of multi-player interactions. Notably, various types of benefit functions are typically considered in public goods games, including linear, saturating, and sigmoid functions. Thus far, the evolutionary dynamics of cooperation in continuous public goods games with these benefit functions remain unknown in structured populations. In this paper, we consider the continuous public goods game in structured populations. By employing the pair approximation approach, we derive the analytical expressions for invasion fitness. Furthermore, we explore the adaptive dynamics of cooperative investments in the game with various benefit functions. First, for the linear public goods game, we find that there is no singular strategy, and the cooperative investments evolve to either the maximum or minimum depending on the benefit-to-cost ratio.  Subsequently, we examine the game with saturating benefit functions and demonstrate the potential existence of an evolutionarily stable strategy (ESS). Additionally, for the game with the sigmoid benefit function, we observe that the evolutionary outcomes are closely related to the threshold value. When the threshold is small, a unique ESS emerges. For intermediate threshold values, both the ESS and repellor singular strategies can coexist. When the threshold value is large, a unique repellor displays. Finally, we perform individual-based simulations to validate our theoretical results.
	\end{abstract}
	
	\maketitle
	
	\textbf{Cooperative behavior is widespread in nature, yet the driving forces for the evolution of cooperation remain poorly understood. While continuous game models have been widely adopted to study the evolutionary dynamics of cooperation, prior research predominantly focuses on pairwise interactions in structured populations, leaving the evolutionary dynamics of continuous multi-players games with group interactions largely unexplored. To address this gap, in this work we investigate the evolutionary dynamics of cooperation in structured populations using continuous public goods games, a paradigm of multi-player games, with varying benefit functions (linear, saturating, and sigmoid). By employing the pair approximation and adaptive dynamics, we derive the conditions for different evolutionary outcomes, showing that cooperation depends on the benefit functions. Our work advances the quantitative understanding of the evolution of cooperation in continuous multi-player games and highlights the pivotal role of nonlinear social returns in stabilizing collective cooperation.}
	
	\section{Introduction}
	Cooperation is a widespread phenomenon ranging from biological systems to human societies ~\cite{axelrod1981evolution,fehr2003nature, nowak2006five, rand2013human}.	However, the emergence and maintenance of cooperative behavior among selfish individuals remains an enduring conundrum. As an alternative approach, evolutionary game theory, a significant mathematical tool, offers a compelling approach to address this challenging issue ~\cite{hofbauer1998evolutionary, smith1982evolution, smith1973logic,Gatenby2005EVOLUTIONARY}. The cooperation conundrum has been studied by employing various game theoretical models. The Prisoner's Dilemma game (PDG), Hawk-Dove game (HDG), Stag Hunt game (SHG), and Snowdrift game (SG) are often employed as the paradigms to study the cooperation problem in a population involving pairwise interactions ~\cite{carlsson199312, doebeli2005models, grafen1979hawk, nakamaru2003can, neugebauer2008fairness, rankin2000strategic, trivers1971evolution}.
	
	However, these models have been traditionally limited to two distinct strategies: cooperation (C) and defection (D). In real-world scenarios, strategies are rarely discretely defined. In contrast, continuous traits may better capture the flexibility of behavior. Hence, the cooperative investment in the game models can be treated as a continuous strategy, varying over a defined range ~\cite{killingback2002continuous, killingback1999variable}.
	In this context, Doebeli et al. analyzed the evolutionary dynamics of the cooperative investments in the continuous snowdrift game ~\cite{doebeli2004evolutionary}. Their study, which focused on well-mixed populations, revealed that cooperative investments in SG can spontaneously diversify, leading to the stable coexistence of both high and low contributions. However, this study concentrated on well-mixed populations, where all individuals interact equally likely with each other. This assumption is an idealization since some individuals interact more frequently than others. In reality, populations are often structured ~\cite{nowak1992evolutionary, szabo1998evolutionary,hauert2004spatial,  killingback2006evolution, Li2022chaos}. Recognizing the significance of population structure, Hauert et al. turned attention to the evolutionary dynamics of cooperative investments in structured populations ~\cite{hauert2021spatial}. Utilizing the adaptive dynamics approach, they found that diversification through evolutionary branching is suppressed compared to well-mixed populations.
	
	Although previous studies have investigated the evolutionary dynamics in continuous games, most of them mainly focus on two-player games. Indeed, group interactions involving multiple individuals rather than pairwise interactions are usually found in real situations. Such multi-player interactions can be effectively described by utilizing \emph{N}-player games ~\cite{chen2017evolutionary, fink1964equilibrium, hamburger1973n, luo2021evolutionary, zheng2007cooperative,Pacheco2012EVOLUTIONARY}. In particular, the public goods game (PGG) is a paradigm of multi-player games ~\cite{chen2012impact, deng2011adaptive,Dong24Chaos, Han15JRSI,hauert2002replicator, hauert2006evolutionary, li2016evolutionary, perc2013Evolutionary, santos2008social, liu2019evolutionary, szolnoki2010reward, zhang2013A}. Thus, the evolution of cooperation within continuous public goods games merits further exploration.
	Notably, when analyzing public goods games, various forms of benefit functions are typically considered. The most common types include linear, saturating, and sigmoid benefit functions ~\cite{chen2012impact, deng2011adaptive, killingback1999variable}. However, the evolutionary dynamics of cooperation in continuous public goods games with these benefit functions remain unknown in structured populations. Therefore, it is worthwhile to investigate the evolutionary dynamics in continuous public goods games with these different forms of benefit functions in structured populations.
	
	In this paper, we thus consider the continuous PGG in the structured population depicted by a regular network. By employing the pair approximation approach, we obtain the invasion fitness and subsequently derive dynamic equations under three different benefit functions, thereby analyzing the evolution of cooperative investments.
	For the PGG with a linear benefit function, our analysis reveals that cooperative investments evolve to either maximum or minimum depending on the benefit-to-cost ratio. We then turn our attention to the game with a saturating benefit function and find a potential singular strategy that serves as an evolutionarily stable strategy (ESS). Next, we analyze the PGG with a sigmoid benefit function. Our observations indicate that when the threshold value in the benefit functions is small, a singular strategy serving as an ESS exists; when the threshold is intermediate, both singular strategies, ESS and repellor, exist; and when the threshold is large, only a unique repellor displays.

	\section{Model}\label{section2}
	
	We consider the continuous PGG in an infinitely structured population represented by a regular network of degree $k$. Individuals are assigned to the nodes of the network and each of them participates in $k+1$ games organized by himself/herself and his/her $k$ neighbors. In each generation, every player organizes a PGG with a group of size $n=k +1$, including himself/herself and his/her neighbors. Following previous researches  ~\cite{doebeli2004evolutionary, hauert2021spatial}, we assume that the trait $x\in[0,1]$ represents the level of cooperative investment that can vary continuously. In particular, extreme cases where $x=1$ and $x=0$ correspond to pure cooperation and pure defection, respectively. We also assume that an individual $i$ contributes an investment $x_i$ to each group he/she participates in.
	
\begin{figure}[htbp]
		\centering
		\includegraphics[scale=0.6]{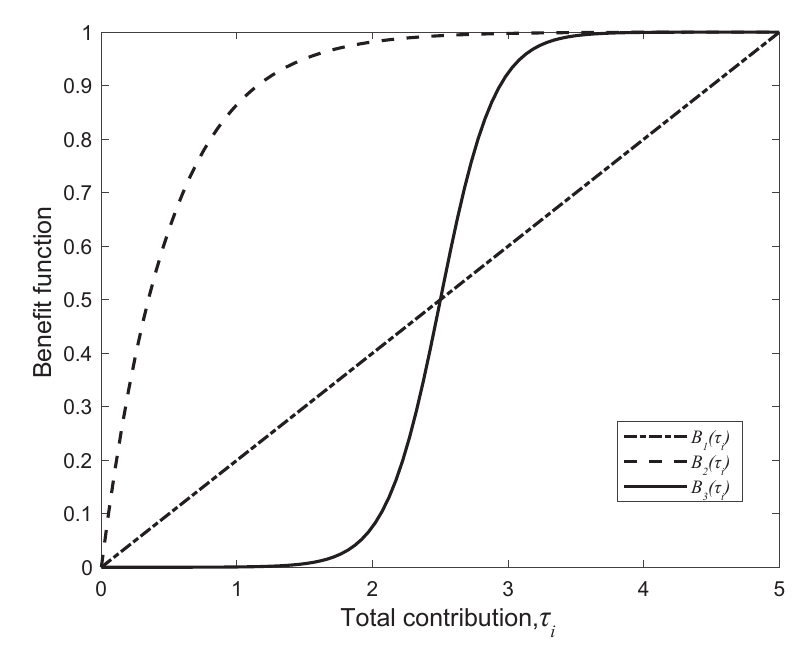}
		\caption{Three kinds of benefit functions of the PGG. The linear benefit function (the dash-dot line), $B_1(\tau_i)=b \tau_i/5$, where $b=1$. The saturating benefit function (the dashed line), $B_2(\tau_i)=b(1-\exp(-\beta_{\text{sat}} \tau_i))$, where $b=1$ and $\beta_{\text{sat}}=2$. The sigmoid benefit function (the solid line) , $B_3(\tau_i)=\frac b{1+\exp(-\beta_{\text{sig}}(\tau_i-T))}$, where $b=1$, $\beta_{\text{sig}}=5$, and $T=5/2$.}
		\label{fig:Fig1}
	\end{figure}

	Subsequently, the payoff for the individual $i$ in the group organized by himself/herself, $G_i$, is given by
	\begin{equation}
		\pi(x_i) = B(\tau_i) - C(x_i),
	\end{equation}
	where $\tau_i=\sum\limits_{x_j\in G_i}{x_j}$ represents the collective contribution in the public pool. Here, $B(\tau_i)$ represents the benefit function depending on the total amount of the produced public good, while $C(x_i)$ represents the cost function associated with the individual investment. We respectively consider linear, saturating, and sigmoid functions for the benefit functions $B(\tau_i)$, by following previous works ~\cite{chen2012impact, deng2011adaptive, killingback1999variable}. Specifically, we consider the following expressions for the benefit function: $B(\tau_i)=b\tau_i/n$ (linear), $B(\tau_i)=b(1-\exp(-\beta_{\text{sat}} \tau_i))$ (saturating), and $B(\tau_i)=\frac b{1+\exp(-\beta_{\text{sig}}(\tau_i-T))}$ (sigmoid). For clarity, these benefit functions are plotted in Fig.~\ref{fig:Fig1}. For the cost function, we adopt the standard linear form $C(x_i)=cx_i$ where $b>c$ , which is widely used in evolutionary public goods games \cite{fehr2000cooperation, hauert2002volunteering}. We stress that the linear cost function is a classical assumption, ensuring the consistency with existing studies~\cite{chen2012impact, deng2011adaptive, killingback1999variable}.
	
	Here, each individual $i$ participates in $k +1$ games where he/she is a member and collects the total payoff from all the involved games, denoted as $\Pi_i$. The payoffs obtained from interactions with their $k$ neighbors determine the birth rates, given by $b_i=\exp(\omega\Pi_i/n)$, where $\omega> 0$ represents the strength of selection ~\cite{nowak2004evolutionary, tarnita2009strategy, traulsen2007pairwise}. This exponential payoff to the birth rate map is designed to ensure that the birth rates are always positive and can be easily converted into probabilities for reproduction.

	After playing the game, the death-birth updating is employed to describe the update process ~\cite{hauert2021spatial, nowak2010evolutionary, ohtsuki2008evolutionary}. To be specific, a random individual is selected to die, and the surrounding neighbours compete for the empty site proportional to their birth rates.
	
	In what follows, through theoretical analysis and individual-based simulations, we respectively investigate the evolutionary dynamics of cooperation in the continuous public goods game with the proposed linear, saturating, and sigmoid functions.
	
	\section{Theoretical analysis}
	The gradual evolution of continuous traits can be effectively described within the framework of adaptive dynamics ~\cite{ dieckmann1996dynamical,geritz1998evolutionarily, landi2013branching, metz1992should, metz1995Adaptive}. We here assume that there are two types of players in the structured population: a mutant type with trait value $y$ and a resident type with trait value $x$ (where $x$ and $y$ respectively represent investment strategies in the continuous game). The pair approximation approach provides a convenient framework to capture the frequency dynamics of strategies, so that the dynamical equation of the trait $x$ can be derived ~\cite{li2014cooperation, matsuda1992statistical, ohtsuki2006replicator, ohtsuki2006simple,sun2023state}.
	Moreover, this approach enables us to determine the central quantity of adaptive dynamics, the invasion fitness $f (x, y)$. This quantity represents the growth rate of a rare mutant $y$ within a monomorphic resident population with trait $x$. It is defined as the per capita growth rate $\dot{p}_m/p_m$ in the limit $p_m\to 0$.
	
	To derive the adaptive dynamics of trait $x$, we consider the selection gradient $D(x )=\frac{\partial f(x,y)}{\partial y}|_{y=x}$. If $D(x ) > 0$, nearby mutants with trait values $y > x$ can successfully invade. Whereas if  $D(x ) < 0$, mutants with $y < x$ will invade. Traits $x^*$ for which $D(x^*)=0$ are known as singular traits. These traits are convergence stable if the Jacobian of the selection gradient, $CS(x^*)=\frac{dD(x)}{dx}|_{x=x^*}$, is negative. In contrast, a trait is evolutionarily stable if the Hessian of the fitness, defined as $ES(x^*)=\frac{\partial^2f(x^*,y)}{\partial y^2}|_{y=x^*}$, is also negative, indicating that the invasion fitness $f (x, y)$ has a (local) maximum at $x^*$.
	
	Considering the two types of stability, there are four potential outcomes when evaluating a singular strategy $x^*$ that satisfies $D(x^*)=0$ ~\cite{doebeli2004evolutionary}.
	
	If the singular trait $x^*$ does not exist, investments evolve either to their maximum or minimum, depending on the sign of the selection gradient.
	
	If $x^*$ exists but is not convergence stable, it acts as a repellor and determines that investments evolve towards either maximum or zero based on their initial investment value $x_0$.
	
	Conversely, if $x^*$ is both convergence stable and evolutionarily stable, it serves as an attractor for stable intermediate investments known as ESS, representing an evolutionary end state.
	
	Finally, if $x^*$ is convergence stable but not evolutionarily stable, it acts as an evolutionary branching point and potential starting point for diversification into coexisting high and low investors.
	
	\subsection{PGG with linear benefit function}
	As the linear benefit function is assumed most frequently in public goods games, we begin by examining the evolution of cooperative investments with a linear benefit function given by $B(\tau_i)=b\tau_i/n$. For the given function, the adaptive dynamics can be expressed as
	\begin{equation}
		\label{eq_adaptivedynamics_sign}
		\dot{x}=\frac{w(n-3)}{(n-2)n}\left(\frac{n+2}{n}b-nc\right)
	\end{equation}
	(see Appendix~\ref{appendix D} for detailed derivations), where $\dot{x}$ represents the time derivative of strategy $x$. The right-hand of Eq.~(\ref{eq_adaptivedynamics_sign}) corresponds to the selection gradient $D(x)$, which symbolically represents the direction of evolutionary change in the strategy space.
	
	It is observed that the selection gradient is reduced to a constant and changes sign at $r^* = (b/c)^*=n^2/(n+2)$. This indicates that there is no singular strategy, leading cooperative investments to evolve toward either pure defection or pure cooperation. To be specific, when $b/c<r^*$, the selection gradient $D(x)$ remains negative for all $x$ in [0,1], suggesting that cooperative investments will decline to the minimum. Conversely, when $b/c > r^*$, only larger mutations can successfully invade, indicating that the trait $x$ evolves towards the maximum. These findings coincide with the evolutionary outcomes observed in the linear PGG under discrete strategies ~\cite{li2016evolutionary}. Additionally, we note that as the group size $n$ increases, the value of $r^*$ also increases. This implies that as the number of individuals in the group grows, it becomes increasingly difficult for cooperation to emerge.
	
	\begin{figure}[H]
		\centering
		\includegraphics[width=8cm]{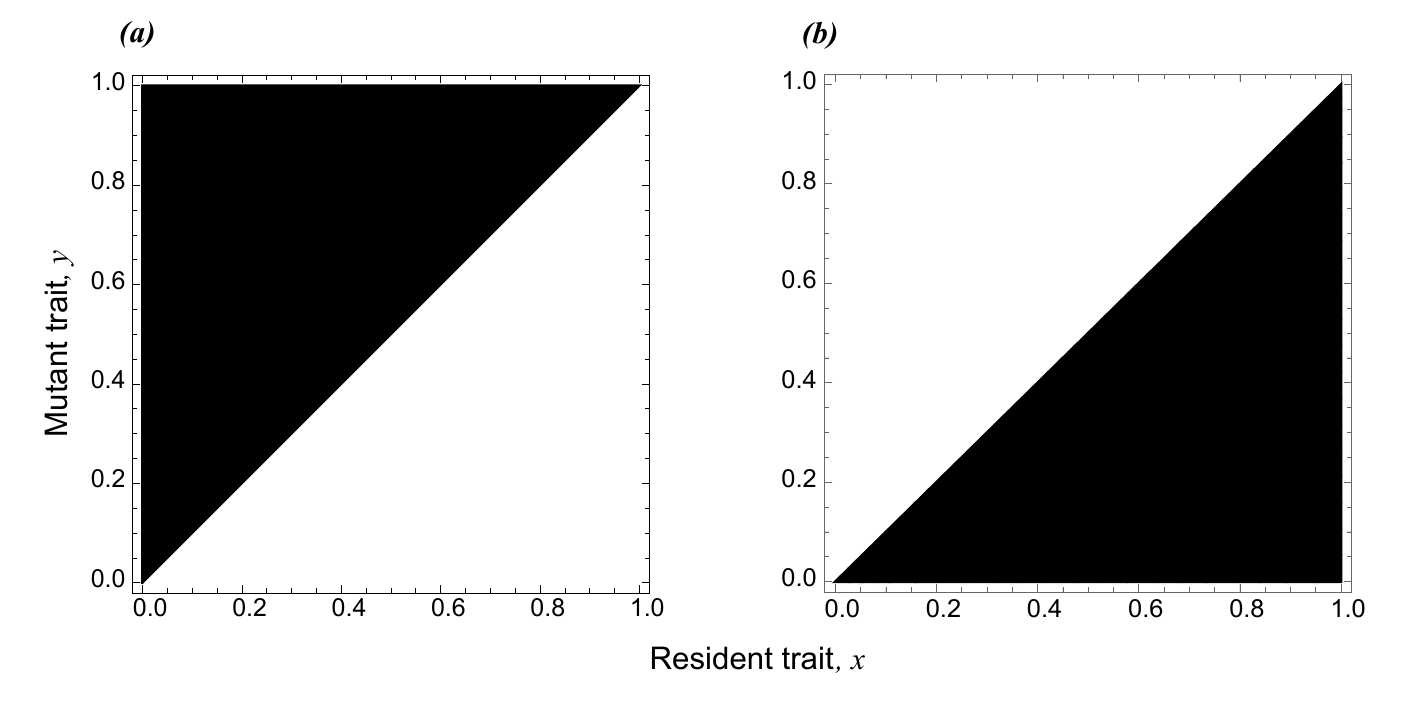}
		\caption{PIP for PGG with linear benefit function $B(\tau_i)=b\tau_i/n$. In the white regions within the separate plots, the invasion fitness is positive, indicating that mutant traits are capable of invading the resident population. Parameter values: $n=5$, $b=2$, $c=1$, and $\omega=1$ in panel ($a$); $n=5$, $b=5$, $c=1$, and $\omega=1$ in panel ($b$).}
		\label{fig:ESS_linear}
	\end{figure}
	
	This analysis can be further elucidated through the pairwise invasion plot (PIP). The PIP visually depicts the regions in the $(x, y)$ space where the mutant strategy $y$ can invade the resident strategy $x$. Specifically, it highlights the regions where the invasion fitness $f(x, y)$ is positive, represented by white regions. From Fig.~\ref{fig:ESS_linear}(a), we observe that $f(x, y)>0$ holds only for $y < x$, indicating that the mutant with lower investment can invade successfully. In contrast, as illustrated in Fig.~\ref{fig:ESS_linear}(b), for $b/c > r^*$, the condition of $f(x, y) > 0$ is only satisfied when $y > x$. It can therefore be inferred that investments will either evolve towards the minimum or maximum depending on the benefit-to-cost ratio when the benefit function is linear.

	\subsection{PGG with saturating benefit function}
	In reality, however, the assumption of linear benefit function may not be hold, as the marginal returns of collective investments often diminish with increasing contributions~\cite{hart1992biological,wilkinson1984reciprocal}. To capture this realistic feature, we next adopt a saturating benefit function defined as $B(\tau_i) = b(1 - \exp(-\beta_{\text{sat}} \tau_i))$. The dynamical equation is given as
	\begin{equation}\label{eq_SG_dynamics_mian}
		\dot{x} = \frac{w(n-3)}{(n-2)n}\left((n+2)b\beta_{\text{sat}} e^{-\beta_{\text{sat}} nx} - nc\right)
	\end{equation}
	(see Appendix~\ref{appendix D} for detailed derivations).
	
\begin{figure}[H]
		\centering
		\includegraphics[width=6cm]{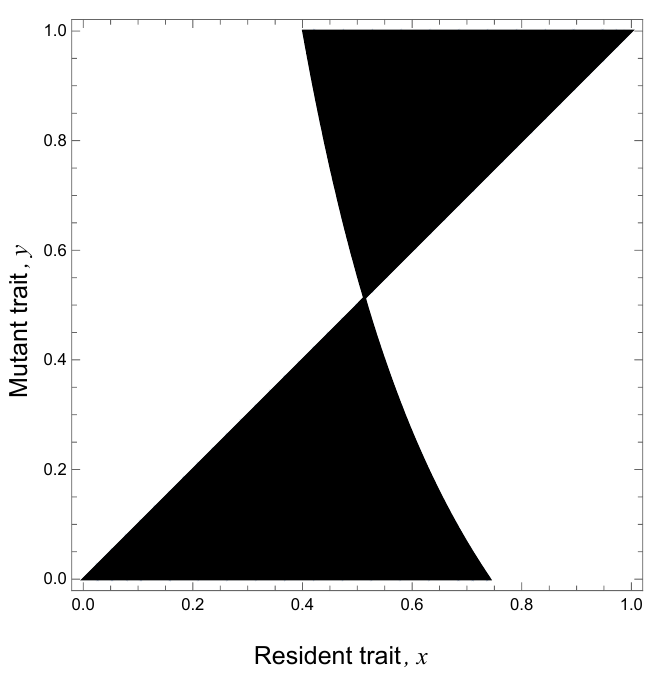}
		\caption{PIP for PGG with saturating benefit function $B (\tau_i)= b (1-\exp(-\beta \tau_i)$. The PIP shows that higher investing mutants can invade for low resident investments, while lower investing mutants can invade high investing residents, and the singular investment level is $x^*=\frac{1}{\beta_{\text{sat}} n}\ln{\left(\frac{(n+2)b\beta_{\text{sat}}}{cn}\right) \approx 0.527811}$. Parameter values: $b=10$, $c=1$, $\beta_{\text{sat}}=1$, and $\omega=1$.}
		\label{fig:ESS_saturating}
	\end{figure}

	Accordingly, we can identify the potential existence of a singular strategy, $x^*=\frac{1}{\beta_{\text{sat}} n}\ln{\left(\frac{(n+2)b\beta_{\text{sat}}}{cn}\right)}$. The presence of this strategy becomes feasible when the benefit-to-cost ratio falls within the range $b/c\in\left(\frac{n}{\beta_{\text{sat}}(n+2)},\frac{ne^{\beta_{\text{sat}} n}}{\beta_{\text{sat}}(n+2)}\right)$. Notably, the lower bound of this range always holds under the assumption of $b > c$, when the parameter $\beta_{\text{sat}}$ is relatively large. Additionally, the upper bound increases significantly due to exponential effects when the parameters $\beta_{\text{sat}}$ and $n$ are relatively large. Consequently, our focus will be on cases where singular strategies exist. By incorporating this strategy into the convergence stability condition and the evolutionary stability condition (see Appendix~\ref{appendix D} for detailed derivations), we determine that the singular strategy is an ESS, satisfying both stability conditions. The two stability criteria respectively ensure the gradual approach to the singular strategy through a series of small evolutionary steps and render a population immune against invasion by any new mutant. According to the PIP shown in Fig.~\ref{fig:ESS_saturating}, it is apparent that only larger mutations can successfully invade when the resident strategy is positioned to the left of the singular strategy, whereas only small mutations can invade when the resident is situated on the right side. Additionally, we observe that as the group size $n$ increases, cooperative investments tend to stabilize at a lower ESS.
	
	\subsection{PGG with sigmoid benefit function}
	While the saturating benefit function captures diminishing returns, many real-world situations exhibit the threshold effects, where only the collective benefit can be shared once contributions surpass a critical level~\cite{szolnoki2010impact}. To model such behavior, we consider a sigmoid benefit function $ B(\tau_i) = \frac{b}{1 + \exp(-\beta_{\text{sig}}(\tau_i - T))}$, where $T > 0$ denotes the threshold and $ \beta_{\text{sig}}$ indicates the steepness of the benefit function. For the given benefit function, the dynamical equation can be expressed as
	\begin{equation}\label{eq_TG_dynamics_main}
		\dot{x}=\frac{w(n-3)}{(n-2)n}\biggl(\frac{(n+2)b\beta_{\text{sig}} e^{-\beta_{\text{sig}}(nx-T)}}{(1+e^{-\beta_{\text{sig}}\left(nx-T\right)})^{2}}-nc\biggr)
	\end{equation}
	(see Appendix~\ref{appendix D} for detailed derivations).
	
	By solving the roots of $D(x^*)=0$, we can see that there exist two potential singular strategies within the interval [0,1], that is,
	\begin{equation}\label{eq_TG_singularpoints_main}
		\begin{aligned}&x_{i}^{*}=\frac{1}{n}(T-\frac{1}{\beta_{\text{sig}}} \ln X_{i})(i=1,2),\\&X_{1}=\frac{b}{c}\cdot\frac{\beta_{\text{sig}}(n+2)}{2n}-1- \sqrt{\left(\frac{b}{c}\cdot\frac{\beta_{\text{sig}}(n+2)}{2n}-1\right)^{2}-1},\\&X_{2}=\frac{b}{c}\cdot\frac{\beta_{\text{sig}}(n+2)}{2n}-1+\sqrt{\left(\frac{b}{c}\cdot\frac{\beta_{\text{sig}}(n+2)}{2n}-1\right)^{2}-1}.
	\end{aligned}\end{equation}
	
	Subsequently, we evaluate the convergence stability and evolutionary stability of the two singular strategies. Specifically, $x_{1}^{*}$ denotes an ESS, while $x_{2}^{*}$ works as a repellor (see \ref{appendix D} for detailed derivations). It is important to note that the two singular strategies $x_{1}^{*}$ and $x_{2}^{*}$ may not always fall within the interval [0, 1].
	
\begin{figure}[H]
		\centering
		\includegraphics[width=8cm]{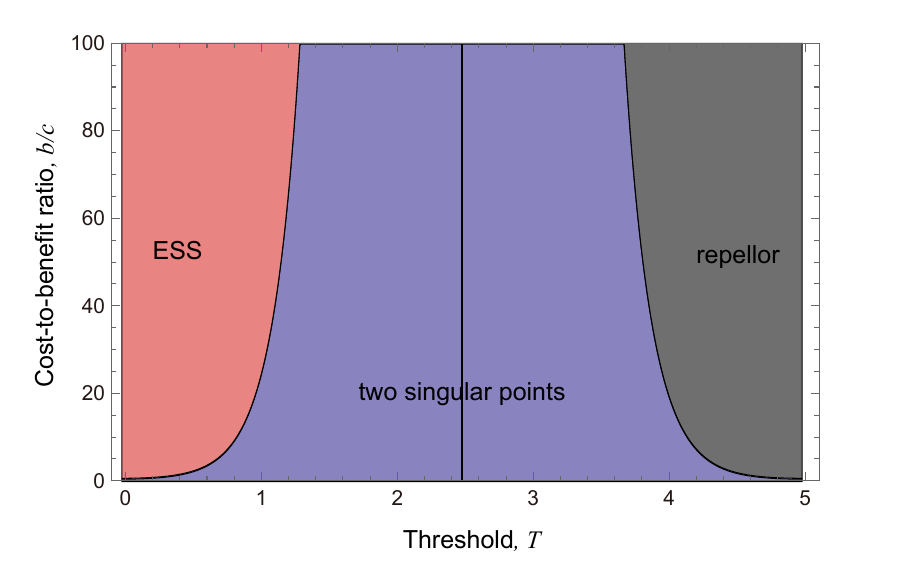}
		\caption{Presence of singular strategies is shown as a function of the benefit-to-cost ratio parameter $b/c$ and threshold parameter $T$ for sigmoid benefit function $B(\tau_i)=\frac b{1+\exp(-\beta_{\text{sig}}(\tau_i-T))}$. Parameter values: $n=5$ and $\beta_{\text{sig}}=5$.}
		\label{fig:singular_sigmoid}
	\end{figure}

	As can be seen in Eq.~(\ref{eq_TG_singularpoints_main}), the magnitude of these two singular strategies is influenced by threshold $T$ and the benefit-to-cost ratio $b/c$. To be specific, when  $T\in[0,n/2)$ and $b/c\in\left(\frac{n(1+e^{\beta_{\text{sig}} T})^2}{\beta_{\text{sig}}(n+2)e^{\beta_{\text{sig}} T}},\\\frac{n(1+e^{\beta_{\text{sig}}(T-n)})^2}{\beta_{\text{sig}}(n+2)e^{\beta_{\text{sig}}(T-n)}}\right)$, a unique ESS denoted as $x^*_1$ exhibits. Furthermore, within the intervals of $T\in[0,n/2)$, $b/c \in \left(\frac{4n}{\beta_{\text{sig}}(n+2)}, \frac{n(1+e^{\beta_{\text{sig}} T})^2}{\beta_{\text{sig}}(n+2)e^{\beta_{\text{sig}} T}}\right)$, and $T\in[n/2,n)$, $b/c \in \left(\frac{4n}{\beta_{\text{sig}}(n+2)}, \frac{n(1+e^{\beta_{\text{sig}}(T-n)})^2}{\beta_{\text{sig}}(n+2)e^{\beta_{\text{sig}}(T-n)}}\right)$, we observe the coexistence of both the ESS $x^*_1$ and the repellor $x^*_2$ within the interval [0,1]. Moreover, within the range of $T\in[n/2,n)$ and
	$b/c\in(\frac{n(1+e^{\beta_{\text{sig}}(T-n)})^2}{\beta_{\text{sig}}(n+2)e^{\beta_{\text{sig}}(T-n)}}, \frac{n(1+e^{\beta_{\text{sig}} T})^2}{\beta_{\text{sig}}(n +2)e^{\beta_{\text{sig}} T}})$, only the repellor $x^*_2$ is present in the interval [0,1]. As illustrated in Fig.~\ref{fig:singular_sigmoid}, we can observe how the presence of singular strategies varies with the benefit-to-cost ratio $b/c$ and the threshold $T$. Specifically, when the threshold $T$ is small, it mainly corresponds to the case of a unique ESS. When the threshold is intermediate, it corresponds to the coexistence of ESS and repellor. When the threshold is large, it corresponds to a unique repellor.
	
	In particular, in the case of $T=n$, if a singular strategy exists, it must be a repellor (see Fig.~\ref{fig:singular_sigmoid}). This finding aligns with the evolutionary outcomes observed in threshold PGG with discrete strategies ~\cite{li2016evolutionary}.
	
	\section{Simulation Results}
	
	\begin{figure*}
		\centering
		\centering
		\includegraphics[width=0.8\textwidth]{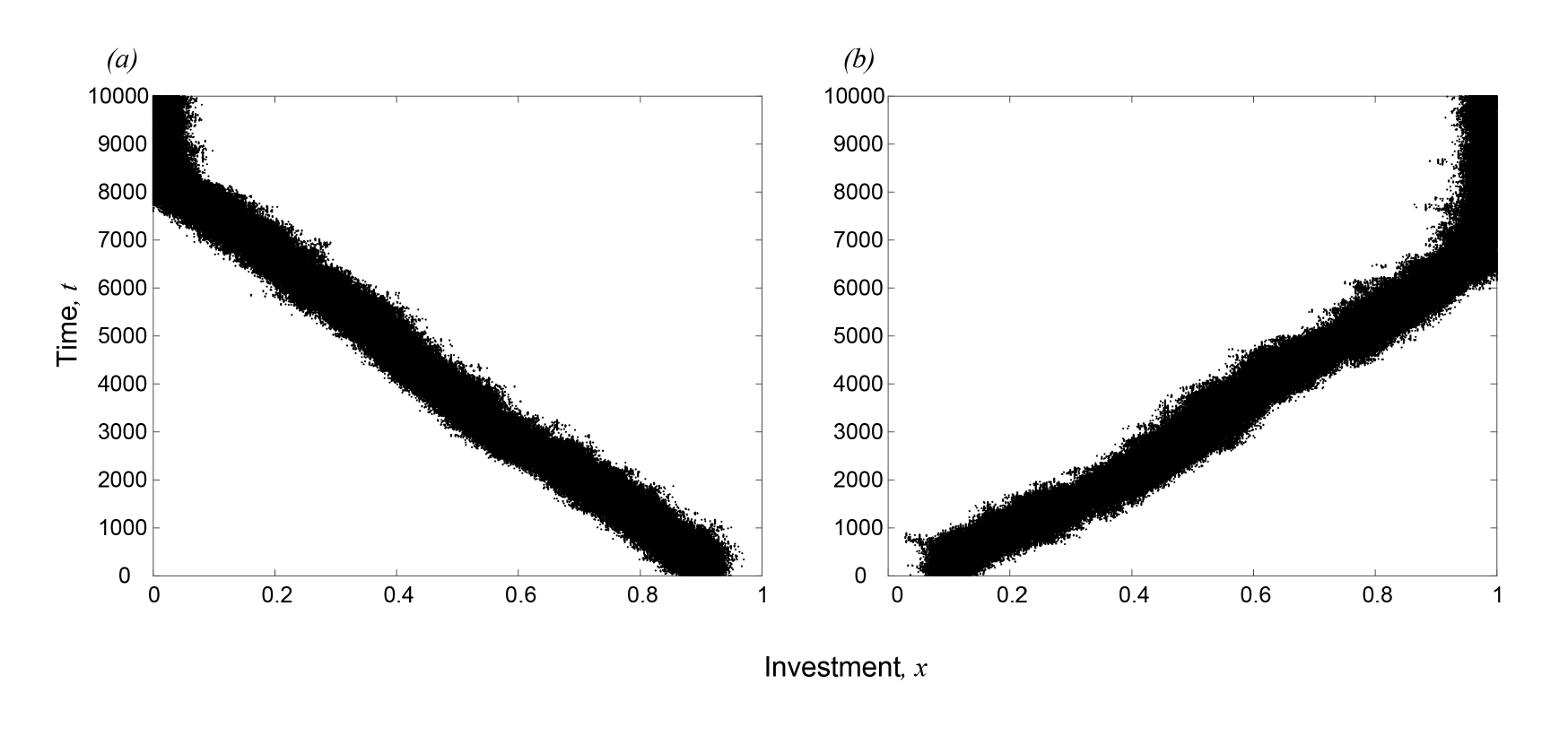}
		\caption{Evolutionary trajectory of the distribution of investments over time for PGG with linear benefit function $B(\tau_i)=b\tau_i/n$. The left panel reveals that the population evolves to full defection when $b/c>r^*$ and the right one shows that the population evolves to full defection when $b/c<r^*$. Parameter values: $b=2$, $c=1$, $x_0=0.9$, and $\omega=1$ in panel (a); $b=5$, $c=1$, $x_0=0.1$, and $\omega=1$ in panel (b).}
		\label{fig:simulation_linear}
	\end{figure*}
	
	\begin{figure*}[!t]
		\centering 
		\includegraphics[scale=0.5]{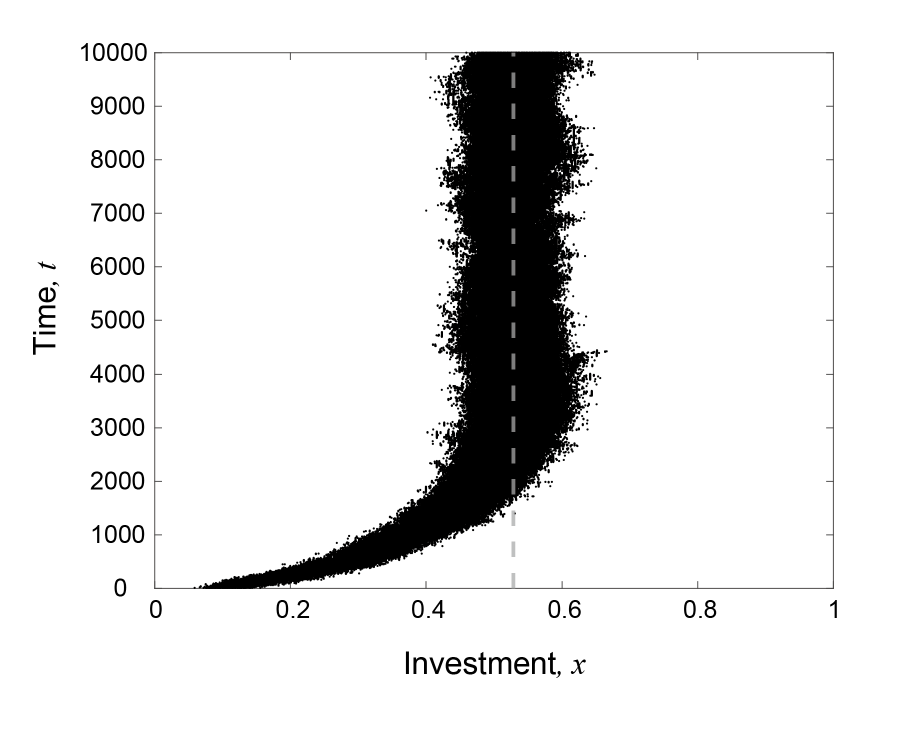}
		\caption{Evolutionary trajectory of the distribution of investments over time for PGG with saturating benefits $B (\tau_i)= b (1-\exp(-\beta_{\text{sig}} \tau_i))$. The gray dashed line represents the theoretically calculated value of the ESS. Parameter values: $b=10$, $c=1$, $\beta_{\text{sig}}=1$, $\omega=1$, and $x_0=0.1$.}
		\label{fig:simulation_saturating}
	\end{figure*}
	
	In this section, we provide individual-based simulations to verify the theoretical analysis presented above. Our simulations are carried out for populations with size $N =10^4$ and run for $10^4$ generations. In structured populations, individuals are arranged on a $100 \times100$ lattice with the von Neumann neighborhood and periodic boundary conditions and they interact with their $k =4$ nearest neighbors. The initial trait value of each individual is randomly assigned from a Gaussian distribution around the initial investment, $x_0$, with standard deviation $\sigma_0 =0.01$. The birth rate of every individual is based on the average payoff of $k+1$ games he/she participates in. For death-birth updating, an individual is randomly selected to die and then all neighbors compete for reproduction. With a probability proportional to his/her birth rate, a neighbor manages to reproduce and place an offspring in the vacant position. Whenever an individual reproduces, the offspring inherits the parental strategy. However, with probability $\mu=0.01$ a mutation occurs and the strategy of the offspring is drawn from a Gaussian distribution around the parental strategy with standard deviation $\sigma_{mut} =0.01$.
	
	We first present the evolutionary trajectory of the distribution of investments over time for the linear benefit function, as depicted in Fig.~ \ref{fig:simulation_linear}. In Fig.~\ref{fig:simulation_linear}(a), we examine the scenario where the benefit-to-cost ratio is below the critical cooperation threshold $r^*$ predicted by adaptive dynamics. The results demonstrate that the investment distribution within the population gradually declines to zero. In contrast, Fig.~\ref{fig:simulation_linear}(b) illustrates the situation where the benefit-to-cost ratio exceeds $r^*$, revealing that cooperative investments evolve from an initially low level to the maximum, ultimately reaching the state of pure cooperation.
	
	Figure \ref{fig:simulation_saturating} shows the evolutionary dynamics of the trait distribution for the saturating benefit function. Based on the theoretical results, it is known that if the singular strategy exists, it must be an ESS. Therefore, in our simulations we consider the benefit-to-cost value at which the existence of singular strategies happens. We find that cooperative investments evolve toward the singular strategy. The distribution of cooperative investment in the population stabilizes around the ESS predicted by adaptive dynamics (the gray dashed vertical line).
	
\begin{figure*}[!t]
		\centering 
		\includegraphics[scale=0.5]{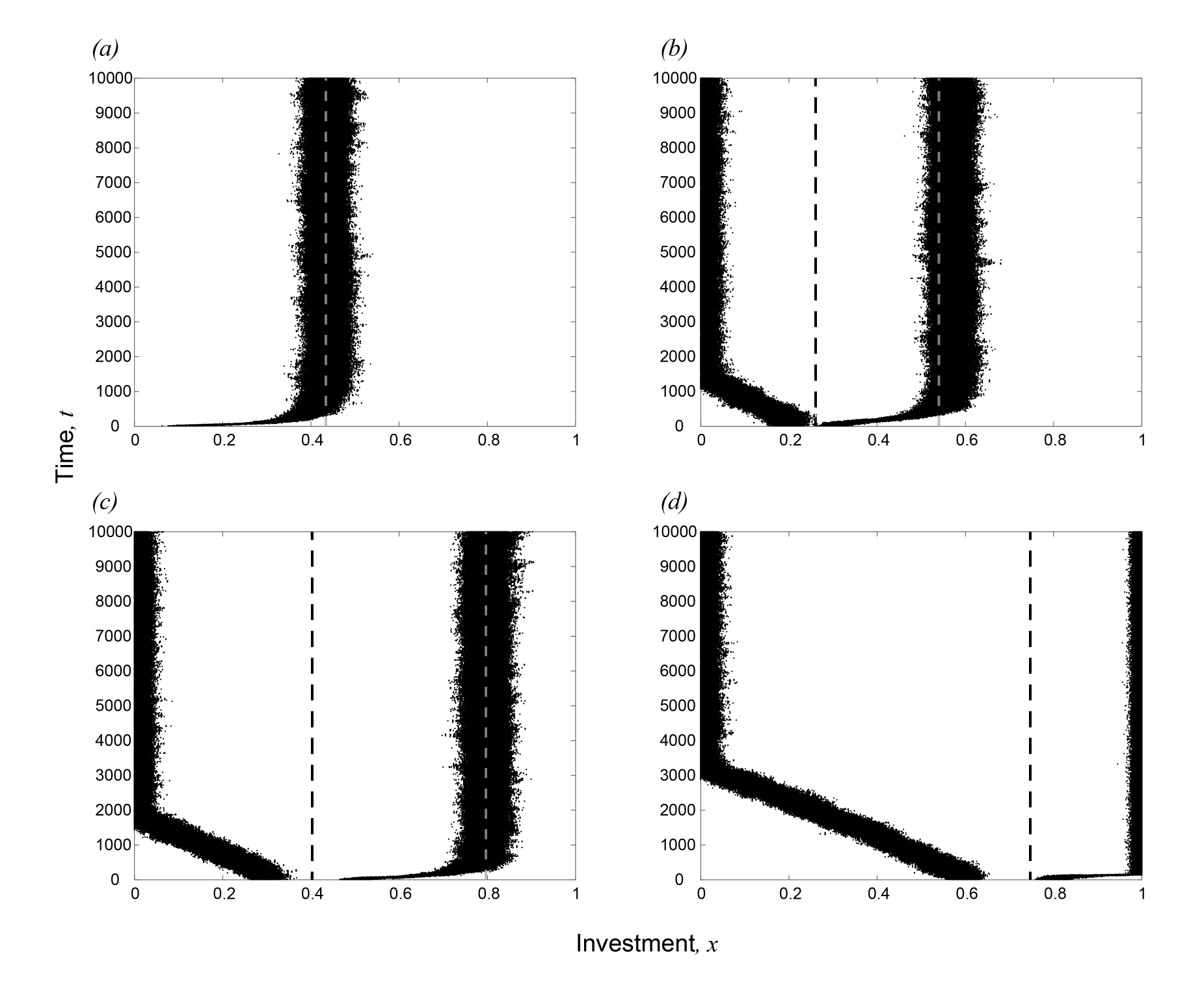}
		\caption{Evolutionary dynamics for PGG with sigmoid benefit function $B(\tau_i)=\frac b{1+\exp(-\beta_{\text{sig}}(\tau_i-T))}$. The top row shows the evolutionary dynamics of the trait distribution for $0<T<n/2$, and the bottom row corresponds to $ n/2< T<n$. The singular strategies (the dashed vertical lines) are indicated where appropriate. The grey lines correspond to the ESS, while the black lines represent the repellor. $(a)$ Evolutionarily stable singular strategy ($x_0=0.1$) exists. $(b)$ Repellor and ESS coexist; depending on the initial conditions, the population either evolves to full defection ($x_0=0.2$) or ESS ($x_0=0.3$). $(c)$ Repellor and ESS coexist; depending on the initial conditions, the population either evolves to full defection ($x_0=0.3$) or ESS ($x_0=0.5$). $(d)$ The existence of repellor happens; depending on the initial conditions, the population either evolves to zero investment ($x_0=0.6$) or full investment ($x_0=0.8$). The grey and black dashed lines represent the theoretically calculated values of ESS and repellor, respectively. Parameter values: $T=1$, $b=50$, and $x^*_1\approx 0.434088$ in panel ($a$); $T=2$, $b=10$, and $x^*_1\approx 0.231228$, $x^*_2\approx 0.568772$ in panel ($b$); $T=3$, $b=20$, and $x^*_1\approx 0.402912$, $x^*_2\approx 0.797088$ in panel ($c$); $T=5$, $b=80$, and $x^*_2\approx 0.747026$ in panel ($d$). Other parameters: $n=5$, $\beta_{\text{sig}}=5$, $c=1$, and $\omega=1$.}
		\label{graph_TG_simulation}
	\end{figure*}

	Figure~\ref{graph_TG_simulation} further illustrates the evolution of the cooperative investment distribution over time when the sigmoid benefit function is utilized. In Fig.~\ref{graph_TG_simulation}(a), cooperative investments evolve towards the singular strategy and stabilize near it, corresponding to the ESS (the gray dashed vertical line). Figure \ref{graph_TG_simulation}(b) and (c) depict the evolutionary trajectories when the threshold values are intermediate. It can be observed that when the initial distribution of cooperative investments is below the theoretically calculated repellor (the black dashed vertical line), cooperative investments evolve away from the repellor. Conversely, when the initial distribution exceeds the value of the repellor, the trait evolves towards the ESS and ultimately stabilizes near it. In the case of a larger threshold $T$ (see Fig.~\ref{graph_TG_simulation}(d)), if the initial investment is positioned to the left of the predicted singular strategy, cooperative investments gradually decrease to zero. In contrast, if it is positioned to the right of the singular strategy, cooperative investments evolve to its maximum value, consistent with the characteristics of the repellor.
	
	\section{Conclusions}
	
	In this paper, we have investigated the evolution of cooperative investments in the spatial PGG with different benefit functions. By employing the pair approximation approach, we have derived the frequency dynamics of the mutant strategy. Subsequently, we have obtained the dynamical equations for these different benefit functions. For the PGG with the linear benefit function, we have found that there are no singular strategies, and the evolutionary outcomes depend significantly on the benefit-to-cost ratio. Concretely, spatial adaptive dynamics predict a benefit-to-cost threshold above which investments reach the maximum but below which they reach the minimum. Subsequently, we have identified a potential ESS when considering the saturating benefit function. Furthermore, for the PGG with the sigmoid benefit function, we have revealed that the evolutionary outcomes are closely associated with the threshold $T$. When the threshold is small, a unique ESS emerges. For the intermediate threshold, both the ESS and repellor singular strategies can coexist. When $T$ is large, it primarily corresponds to a unique repellor.
	
	This work continues along the lines of previous investigations considering continuous games with different benefit functions ~\cite{hauert2021spatial}, but it specifically focuses on the PGG with group interactions rather than pairwise interactions. We have derived the mathematical conditions for the emergence of cooperation. Notably, we have observed that cooperation emerges in PGG at a lower benefit-to-cost threshold compared to the PDG. Specifically, under the linear benefit function, cooperation in PGG emerges when the benefit-to-cost ratio satisfies \( b/c > \frac{(k+1)^2}{k+3} \),  a condition that is less stringent than that of the PDG, where cooperation emerges at \( b/c > k \). For the saturating benefit function, the condition for cooperation to emerge is \( b/c > \frac{(k+1)}{\beta_{\text{sat}}(k+3)} \), which is also weaker than the cooperation condition in the PDG, where $b/c >k/\beta_{\text{sat}}$.
	
	In comparison to Ref.~\cite{li2016evolutionary}, which examined the evolution of cooperation in the spatial public goods game with two discrete strategies, we investigate the evolution of continuous investments. For linear benefit functions, we find that strategies eventually evolve toward either full cooperation or full defection in both the discrete and continuous scenarios. Notably, the range of benefit-to-cost ratios corresponding to these two evolutionary outcomes coincides exactly in both scenarios. Indeed, Ref.~\cite{li2016evolutionary} considered the evolution of cooperation when the benefit function is a step-like function. For the convenience of theoretical analysis, the authors analyzed the evolutionary dynamics when the parameter threshold is set to the maximum $T=n$ and identified the potential existence of a repellor. This evolutionary outcome aligns with our findings in PGG with sigmoid benefit functions. The discrepancy in benefit-to-cost ratios associated with the repellor in continuous and discrete scenarios may be attributed to the use of distinct benefit functions. Remarkably, we consider a generalized sigmoid benefit function in our study. In addition, we explore how the varying threshold $T$ can impact evolutionary outcomes. Our work thus further enriches the knowledge of the evolutionary dynamics of PGG with different benefit functions.
	
In this work, we have focused on the evolutionary dynamics of continuous public goods games in regular networks. In future work, it would be interesting to study the evolutionary dynamics in heterogeneous networks~\cite{allen2017evolutionary,cimpeanu2022artificial}. Furthermore, it is also worthwhile to study the evolutionary dynamics of cooperation when incentives are integrated into the continuous PGG we considered~\cite{wang2019exploring,cimpeanu2023does}.
	
	\appendix
	\section{Calculation details under the pair approximation approach}
	\setcounter{equation}{0}
	\renewcommand{\theequation}{A\arabic{equation}}
	
	For continuous games in spatially structured populations, the frequency dynamics of two types, the resident and the mutant, can be modeled using the pair approximation approach ~\cite{li2014cooperation, matsuda1992statistical, ohtsuki2006replicator, ohtsuki2006simple,sun2023state}. Correspondingly, we use the notations $P_X$ and $P_{XY}$ to respectively represent the frequencies of \emph{X} resident and \emph{XY} pairs in the population. Let $q_{X|Y}$ denote the conditional probability to find an individual with strategy \emph{X} given that the adjacent node is occupied by a neighbor with strategy \emph{Y}. Let both \emph{X} and \emph{Y} stand for \emph{m} and \emph{r}, and based on the above-mentioned descriptions we have
	\begin{equation}
		P_m+P_r=1,
	\end{equation}
	\begin{equation}
		P_{mr}=P_{rm},
	\end{equation}
	\begin{equation}
		P_{mr}=P_r q_{m|r},
	\end{equation}
	and
	\begin{equation}
		q_{m|m}+q_{r|m}=1.
	\end{equation}
	Note that the whole system can be fully described by using only two variables $p_m$ and $q_{m|m}$. Moreover, the respective rates of change depend on the microscopic updating procedure.
	In the following, we calculate the dynamical changes of the frequency of mutant $p_m$ for death-birth updating.
	
	\subsection{Updating a resident}\label{appendix B}
	
	For the two types: a mutant type with trait value $y$ and a resident type with trait value $x$, the collected contribution in the public pool organized by individual $i$ equals $\tau_i(j)=jy+(n-j)x$, where $j$ represents the number of $y$-strategists in the group.
	Correspondingly, the payoff of the mutant and resident in one interaction group can be respectively given by
	\begin{equation}
		\pi_m(j) = B(jy+(n-j)x) - C(y),
	\end{equation}
	and
	\begin{equation}
		\pi_r(j) = B(jy+(n-j)x) - C(x),
	\end{equation}
	where $n=k+1$ represents the group size.
	
	We further consider that a resident is replaced by a neighboring mutant in the regular network. We assume that there are $j$ mutants among the $k$ nearest neighbors. By considering the possible configurations around the resident, we can estimate its mutant and resident neighbors' payoff values. Accordingly, the total payoff of the mutant player, who is the neighbor of the selected resident individual, can be given as
	\begin{equation}
		\begin{aligned}
			\Pi_{m}^{r}(j) &= \pi_m(j)\\
			&+\sum_{i=0}^{k-1}\binom{k-1}iq_{m|m}^i(1-q_{m|m})^{k-1-i}(\pi_m(i+1)  \\
			&+i\sum_{t=0}^{k-1}\binom{k-1}tq_{m|m}^t(1-q_{m|m})^{k-1-t}\pi_m(t+2) \\
			&+(k-1-i)\sum_{t=0}^{k-1}\binom{k-1}tq_{m|r}^t(1-q_{m|r})^{k-1-t}\pi_m(t+1)).
		\end{aligned}
	\end{equation}
	
	The payoff of his/her resident neighbor can be given as
	\begin{equation}
		\begin{aligned}
			\Pi_{r}^{r}(j) &=\pi_r(j)\\
			&+\sum_{i=0}^{k-1}\binom{k-1}iq_{m|r}^i(1-q_{m|r})^{k-1-i}(\pi_r(i)  \\
			&+i\sum_{t=0}^{k-1}\binom{k-1}tq_{m|m}^t(1-q_{m|m})^{k-1-t}\pi_r(t+1) \\
			&+(k-1-i)\sum_{t=0}^{k-1}\binom{k-1}tq_{m|r}^t(1-q_{m|r})^{k-1-t}\pi_r(t)).
		\end{aligned}
	\end{equation}
	Consequently, the birth rates of the neighboring mutant and resident denote as $b_{m}^{r}(j)= e^{\omega\Pi_{m}^{r}(j)/(k+1)}$ and $b_{r}^{r}(j)= e^{\omega\Pi_{r}^{r}(j)/(k+1)}$, respectively. The birth rate is proportional to the probability of taking over an empty site for which a given mutant or resident individual competes.
	
	For death-birth updating ~\cite{hauert2021spatial,ohtsuki2008evolutionary,nowak2010evolutionary}, the frequency of mutants and mutant-mutant pairs increases whenever a resident dies and a mutant neighbor repopulates the vacated site. For a resident with $j$ mutant neighbors, this happens with probability
	\begin{equation}
		\begin{aligned}\label{eq_Tplus}
			T^+(j)&=(1-p_m)\binom{k}jq_{m|r}^jq_{r|r}^{k-j}\frac{jb_m^r(j)}{jb_m^r(j)+(k-j)b_r^r(j)}.\\
		\end{aligned}
	\end{equation}
	
	\subsection{Updating a mutant}\label{appendix C}
	
	We here consider that a mutant is selected and replaced by a neighboring resident. We also assume that there are $j$ mutants among the $k$ nearest neighbors. By considering the possible configurations around the mutant, we can estimate his/her resident and mutant neighbors' payoff values. Accordingly, the total payoff of the resident player, who is the neighbor of the selected mutant individual, can be given as
	\begin{equation}
		\begin{aligned}
			\Pi_{r}^{m}(j) &=\pi_{r}(j+1)\\
			&+\sum_{i=0}^{k-1}\binom{k-1}iq_{m|r}^{i}(1-q_{m|r})^{k-1-i}(\pi_{r}(i+1)\\
			&+i\sum_{t=0}^{k-1}\binom{k-1}tq_{m|m}^t(1-q_{m|m})^{k-1-t}\pi_r(t+1)\\
			&+(k-1-i)\sum_{t=0}^{k-1}\binom{k-1}tq_{m|r}^t(1-q_{m|r})^{k-1-t}\pi_r(t)).
	\end{aligned}\end{equation}
	
	The payoff of his mutant neighbor can be given as
	\begin{equation}
		\begin{aligned}
			\Pi_{m}^{m}(j) &=\pi_m(j+1)\\
			&+\sum_{i=0}^{k-1}\binom{k-1}iq_{m|m}^i(1-q_{m|m})^{k-1-i}(\pi_m(i+2)  \\
			&+i\sum_{t=0}^{k-1}\binom{k-1}tq_{m|m}^t(1-q_{m|m})^{k-1-t}\pi_m(t+2) \\
			&+(k-1-i)\sum_{t=0}^{k-1}\binom{k-1}tq_{m|r}^t(1-q_{m|r})^{k-1-t}\pi_m(t+1)).
		\end{aligned}
	\end{equation}
	
	Similarly, the frequency of mutants and mutant-mutant pairs decreases if a mutant dies and one of his/her resident neighbors reproduces. For a mutant with $j$ mutant neighbors, this happens with probability
	\begin{equation}
		\begin{aligned}
			T^-(j)&=p_m\binom{k}jq_{m|m}^jq_{r|m}^{k-j}\frac{(k-j)b_r^m(j)}{(k-j)b_r^m(j)+jb_m^m(j)}.\\
		\end{aligned}
	\end{equation}
	
	\section{Adaptive dynamics of continuous PGG}\label{appendix D}
	\setcounter{equation}{0}
	\renewcommand{\theequation}{B\arabic{equation}}
	Based on the above-mentioned description, $T^+$ and $T^-$ respectively represent the probability that $p_m$ will increase or decrease by $1/N$. We assume that each invasion step takes place in one unit of time $1/N$ ~\cite{ohtsuki2006replicator, sun2023state}. Hence the derivative of $p_m$ is given by
	\begin{equation}
		\dot{p}_m=\sum_{j=0}^{k}(T^{+}(j)-T^{-}(j)).
	\end{equation}
	
	To derive the change rate in $q_{m|m}$, it helps to start with the changes in $p_{mm}$. First, consider a resident with $j$ mutant neighbors who are successfully replaced by a mutant. This happens with probability $T^+(j)$ and increases the number of $mm$-pairs by $j$ or, equivalently, their frequency $p_{mm}$ by $2j/(Nk)$, where $Nk/2$ is the total number of undirected links in a regular graph of size $N$ and degree $k$. Similarly, with probability $T^-(j)$ a mutant with $j$ mutant neighbors will be replaced by a resident, reducing the frequency $p_{mm}$ by $2j/(Nk)$. So the change rate of $p_{mm}$ is
	\begin{equation}
		\dot{p}_{mm}=\sum_{i=0}^k(T^+(j)-T^-(j))\frac{2j}k.
	\end{equation}
	
	Finally, using $\dot{q}_{m|m}=(\dot{p}_{mm}-q_{m|m}\dot{p}_m)/p_m$ results in
	\begin{equation}
		\dot{q}_{m|m}=\frac{1}{p_{m}}\sum_{j=0}^{k}(T^{+}(j)-T^{-}(j))\left(\frac{2j}{k}-q_{m|m}\right).
	\end{equation}
	
	In the limit $p_{m}\rightarrow0$, the sum over $T^+(j)$ in Eq.~(\ref{eq_Tplus}) and divided by $p_{m}$ is reduced to
	\begin{equation}
		\begin{gathered}
			\lim_{p_{m}\to0}\frac{1}{p_{m}}\sum_{j=0}^{k}T^{+}(j)
			\begin{aligned}=\frac{k(1-q_{m|m})b_{m}^{r}(1)}{b_{m}^{r}(1)+(k-1)b_{r}^{r}(1)}.\end{aligned}
		\end{gathered}
	\end{equation}
	
	Thus the invasion fitness becomes
	\begin{equation}\label{eq_fitness}
		\begin{aligned}
			f(x,y)&=\lim_{p_{m}\to0}\frac{\dot{p}_{m}}{p_{m}}\\
			&=\lim_{p_{m}\to0}\frac{1}{p_{m}}\sum_{j=0}^{k}(T^{+}(j)-T^{-}(j)) \\ &=\frac{k(1-q_{m|m})b_{m}^{r}(1)}{b_{m}^{r}(1)+(k-1)b_{r}^{r}(1)}\\
			&-\sum_{j=0}^{k}\binom{k}jq_{m|m}^{j}(1-q_{m|m})^{k-j}\\
			&\times\frac{(k-j)b_{r}^{m}(j)}{jb_{m}^{m}(j)+(k-j)b_{r}^{m}(j)},
		\end{aligned}
	\end{equation}
	and the change rate of $q_{m|m}$ can be simplified as
	\begin{equation}\label{eq_qmm*}
		\begin{aligned}
			\dot{q}_{m|m} &= \frac{k(1 - q_{m|m})b_{m}^{r}(1)}{b_{m}^{r}(1)+(k - 1)b_{r}^{r}(1)}\left(\frac{2}{k}-q_{m|m}\right)\\
			&-\sum_{j = 0}^{k}\binom{k}{j}q_{m|m}^{j}q_{r|m}^{k - j}\frac{(k - j)b_{r}^{m}(j)}{jb_{m}^{m}(j)+(k - j)b_{r}^{m}(j)}\\
			&\times\left(\frac{2j}{k}-q_{m|m}\right).
		\end{aligned}
	\end{equation}
	
	The local frequency $q_{m|m}$ equilibrates much faster than the global frequency $p_m$ when mutants are rare. This leads to a convenient separation of time scales, allowing us to use the equilibrium $q^*_{m|m}$ of Eq.~(\ref{eq_qmm*}) to calculate invasion fitness
	\begin{equation}
		\begin{aligned}
			f(x,y)&=\frac{k(1-q_{m|m}^*)b_m^r(1)}{b_m^r(1)+(k-1)b_r^r(1)}\\
			&-\sum_{j=0}^k\binom{k}jq_{m|m}^{*j}(1-q_{m|m}^*)^{k-j}\frac{(k-j)b_r^m(j)}{jb_m^m(j)+(k-j)b_r^m(j)}.
		\end{aligned}
	\end{equation}
	
	To calculate $q^*_{m|m}$, we first note that in the limit of rare mutants, $p_{m}\rightarrow0$, and for mutant traits $y$ close to the resident trait $x$, a Taylor expansion of the right-hand side of Eq.~(\ref{eq_qmm*}) up to the first order becomes
	\begin{equation}\label{eq_qmm*taylor}
		\begin{aligned}
			\dot{q}_{m|m}&=\frac{2(1-q_{m|m})(1-(k-1)q_{m|m})}{k}\\
			&+(y-x) \frac{\omega (k-1)(1-q_{m|m})}{k(k+1)} \\
			&\times \biggl((1-\frac{4}{k})q_{m|m}^2+(\frac{2}{k}-1)q_{m|m}+\frac{2}{k}\biggr)\\ &\times\biggl(((k-1)^2q_{m|m}^2+2(k-1)q_{m|m}+k)B'((k+1)x)\\
			&-(k+1)C'(x)\biggr)\\
			&+O((y-x)^2).
		\end{aligned}
	\end{equation}
	
	To simplify the analysis, we can obtain the Taylor expansion of $q_{m|m}^*$ in $y$ around $x$ as
	\begin{equation}
		q_{m|m}^*=q_{m|m}^{*(0)}+q_{m|m}^{*(1)}(y-x)+O((y-x)^2).
	\end{equation}
	The zero order approximation of $q_{m|m}^*$ in $y$ near $x$ can be obtained by solving for the roots of Eq.~(\ref{eq_qmm*taylor}) for $y=x$ is
	\begin{equation}
		q_{m|m}^{*(0)}=\frac{1}{k-1}.
	\end{equation}
	
	Next, the first-order approximation of $q_{m|m}^*$ is obtained by implicitly differentiating Eq.~(\ref{eq_qmm*taylor}) to $y$ (noting that $q_{m|m}$ is a function of $y$) and evaluating at $y=x$. Setting the expression to zero gives an equation for the zeroth and first-order coefficients, $q_{m|m}^{*(0)}$ and $q_{m|m}^{*(1)}$, of the Taylor expansion at the equilibrium $q_{m|m}^{*}$ for $y$ near $x$:

		\begin{equation}
		\begin{aligned}
			& -\frac{2q_{m|m}^{*(1)}\left(1 - (k - 1)q_{m|m}^{*(0)}\right)}{k} \\
			& +\frac{2(1 - q_{m|m}^{*(0)})(1 - k)q_{m|m}^{*(1)}}{k} \\
			& +\frac{\omega (k - 1)(1 - q_{m|m}^{*(0)})}{k(k + 1)}\\
			&\times\biggl((1-\frac{4}{k}){q_{m|m}^{*(0)}}^2+(\frac{2}{k}-1)q_{m|m}^{*(0)}+\frac{2}{k}\biggr)\\
			& \times\biggl(((k - 1)^{2}{q_{m|m}^{*(0)}}^2 + 2(k - 1)q_{m|m}^{*(0)} + k)B^{\prime}((k + 1)x) \\
			& - (k + 1)C^{\prime}(x) \biggr) \\
			& = 0.
		\end{aligned}
	\end{equation}

	Then the first order coefficient $q_{m|m}^{*(1)}$ is given as
	\begin{equation}
		q_{m|m}^{*(1)}=\frac{\omega(k^2-4)}{2(k-1)^2k(k+1)}((k+3)B^{\prime}((k+1)x)-(k+1)C^{\prime}(x)).
	\end{equation}
	Assembling all the pieces finally leads to the first-order expansion in $y$ of the local pair density equilibrium, $q_{m|m}^*$, around $x$ as
	\begin{equation}
		\begin{aligned}
			q_{m|m}^{*}&=\frac{1}{k-1}+(y-x)\frac{\omega(k^2-4)}{2(k-1)^2k(k+1)}\\
			&\times((k+3)B^{\prime}((k+1)x)-(k+1)C^{\prime}(x)).
		\end{aligned}
	\end{equation}
	
	For consistency of notation, we use $x$ to denote the cooperative investment of the mutant, so the dynamical equation can be written as
	\begin{equation}\dot{x}=\frac{\partial f(x,y)}{\partial y}|_{y=x}.\end{equation}
	As a result, the dynamical equation of the continuous trait $x$ is given by
	\begin{equation}\label{eq_dynamics}
		\dot{x}=\frac{\omega(k-2)}{(k-1)(k+1)}\left((k+3)B^{\prime}((k+1)x)-(k+1)C^{\prime}(x)\right).
	\end{equation}
	Since we are considering the group size $n=k+1$, it follows that
	\begin{equation}\dot{x}=\frac{\omega(n-3)}{(n-2)n}\left((n+2)B^{\prime}(nx)-nC^{\prime}(x)\right).\end{equation}
	
	The solutions of the dynamical equation $\dot{x}=D(x)$ are called singular strategies. If a singular strategy $x^*$ exists, it is convergent stable if
	\begin{equation}\label{eq_CS}
		\begin{aligned}
			CS(x^*)&=\frac{dD(x)}{dx}|_{x=x^*}\\ &=\frac{\omega(n-3)}{(n-2)n}\left((n+2)B''(nx^*)-nC''(x^*)\right) \\&<0,
		\end{aligned}
	\end{equation}
	and it is evolutionarily stable if
	\begin{equation}
		\begin{aligned}
			ES(x^*)&=\frac{\partial^2f(x^*,y)}{\partial y^2}|_{y=x^*}\\ &=\frac{w(n-3)}{(n-2)n}\Big((n^3+9n^2-31n+19)B^{\prime\prime}(nx^*)\\
			&-(n-1)nC^{\prime\prime}(nx^*)\Big) \\&<0.
		\end{aligned}
		\label{eq_ES}
	\end{equation}

	There are several salient regimes for the evolutionary dynamics based on the stability of singular strategies ~\cite{doebeli2004evolutionary}:
	1) In cases where $x^*$ does not exist, investments evolve towards either their maximum or minimum values, determined by the sign of the selection gradient.
	2) If $x^*$ exists but is not convergence stable, it acts as a repellor. The evolutionary outcome is influenced by the initial investment $x_0$: for $x_0 > x^*$, investments evolve towards the maximum; while for $x_0 < x^*$, investments tend towards zero.
	3) When $x^*$ is both convergence stable and evolutionarily stable, it serves as an attractor for stable intermediate investments, representing the evolutionary end state.
	4) If $x^*$ is convergence stable but not evolutionarily stable, it signifies an evolutionary branching point, potentially leading to diversification into coexisting high and low investors.
	
	\subsection{Adaptive dynamics of PGG with linear benefit function}
	We first investigate the adaptive dynamics under the linear benefit function $B(\tau_i)=b \tau_i/n$. Referring to Eq.~(\ref{eq_dynamics}), we obtain
	
	\begin{equation}
		\dot{x} = \frac{w(n-3)}{(n-2)n}\left(\frac{n+2}{n}b - nc\right) \equiv D(x).
	\end{equation}
	The selection gradient $D(x)$ has no root in the interval [0, 1], and remains negative when $b/c < \frac{n^2}{n+2}$ and positive when $b/c > \frac{n^2}{n+2}$.
	
	\subsection{Adaptive dynamics of PGG with saturating benefit function}
	When the benefit function is saturating, $B(\tau_i) = b(1-e^{-\beta_{\text{sat}} \tau_i})$, the dynamical equation is
	
	\begin{equation}
		\dot{x} = \frac{w(n-3)}{(n-2)n}\left((n+2)b\beta_{\text{sat}} e^{-\beta_{\text{sat}} nx} - nc\right).
	\end{equation}
	For $b/c < \frac{n}{\beta_{\text{sat}}(n+2)}$, $D(x)$ remains negative for all $x \in [0, 1]$, implying that the strategy $x$ evolves towards zero over time. Similarly, when $b/c > \frac{ne^{\beta_{\text{sat}} n}}{\beta_{\text{sat}}(n+2)}$, $D(x) > 0$ for all $x \in [0, 1]$ and the trait $x$ increases to the maximum. In the case where $\frac{n}{\beta_{\text{sat}}(n+2)} < b/c < \frac{ne^{\beta_{\text{sat}} n}}{\beta_{\text{sat}}(n+2)}$, there exists an internal root of $D(x^*)=0$ and the strategy $x^*$ satisfies both the convergence stability condition $CS(x^*) < 0$ and evolutionary stability condition $ES(x^*) < 0$. Therefore, the unique strategy $x^* = \frac{1}{\beta_{\text{sat}} n} \log{(\frac{(n+2)b\beta_{\text{sat}}}{cn})}$ is an ESS.

	\subsection{Adaptive dynamics of PGG with sigmoid benefit function}
	For PGG with the sigmoid benefit function represented by $B(\tau)=\frac b{1+\exp(-\beta_{\text{sig}}(\tau-T))}$, by utilizing Eq.~(\ref{eq_dynamics}) we derive the dynamical equation
	\begin{equation}\dot{x}=\frac{w(n-3)}{(n-2)n}\biggl(\frac{b\beta_{\text{sig}}(n+2) e^{-\beta_{\text{sig}}(nx-T)}}{(1+e^{-\beta_{\text{sig}}\left(nx-T\right)})^{2}}- nc\biggr). \end{equation}
	It can be seen that there are two potential roots of $D(x)$ within the interval [0, 1]:
	\begin{equation}\label{eq_TGroot}
		\begin{aligned}&x_{i}^{*}=\frac{1}{n}(T-\frac{1}{\beta_{\text{sig}}} \ln X_{i})(i=1,2),\\&X_{1}=\frac{b}{c}\cdot\frac{\beta_{\text{sig}}(n+2)}{2n}-1- \sqrt{\left(\frac{b}{c}\cdot\frac{\beta_{\text{sig}}(n+2)}{2n}-1\right)^{2}-1},\\&X_{2}=\frac{b}{c}\cdot\frac{\beta_{\text{sig}}(n+2)}{2n}-1+\sqrt{\left(\frac{b}{c}\cdot\frac{\beta_{\text{sig}}(n+2)}{2n}-1\right)^{2}-1}.\end{aligned}
	\end{equation}
	By referring to Eqs.~(\ref{eq_CS}) and (\ref{eq_ES}), the singular strategies are convergence stable if
	
	\begin{equation}\label{eq_TGCS}
		\begin{aligned}
			CS(x^*)&=\frac{\omega b\beta_{\text{sig}}^2(n - 3)(n + 2)e^{-\beta_{\text{sig}}\left(nx^*-T\right)}}{(n - 2)(1 + e^{ -\beta_{\text{sig}}\left(nx^*-T\right)})^3} \\
			&\times (e^{-\beta_{\text{sig}}\left(nx^*-T\right)}-1)<0,
		\end{aligned}
	\end{equation}
	and evolutionarily stable if
	
	\begin{equation}\label{eq_TGES}
		\begin{aligned}
			ES(x^*)&=\frac{\omega b\beta_{\text{sig}}^2(n - 3)(n^3 + 9n^2 - 31n + 19)e^{-\beta_{\text{sig}}\left(nx^*-T\right)}}{(n - 2)^2(n - 1)n(1 + e^{-\beta_{\text{sig}}\left(nx^*-T\right)})^3} \\
			&\times(e^{-\beta_{\text{sig}}\left(nx^ *-T\right)}-1)<0.
		\end{aligned}
	\end{equation}
	
	Through Eq.~(\ref{eq_TGroot}), it is apparent that there exist no roots of the selection gradient when $\frac{b}{c} \cdot \frac{\beta_{\text{sig}}(n+2)}{2n} - 1 < 1$. In contrast, in cases where this condition is not met, there are two roots, although not necessarily falling within the interval [0, 1]. It can be seen that $0 < X_1 < 1 < X_2$. Consequently, $x^{*}_1 > x^{*}_2$, $nx^{*}_1 - T > 0$, and $nx^{*}_2 - T < 0$. Hence, $x^{*}_2$ fails to meet the convergence stability condition presented in Eq.~(\ref{eq_TGCS}) and acts as a repellor. Conversely, $x^{*}_1$ is an ESS that satisfies both the convergence stability condition Eq.~(\ref{eq_TGCS}) and the evolutionary stability condition Eq.~(\ref{eq_TGES}). Next, we present the existence of singular strategies for diverse threshold $T$ and benefit-to-cost ratio values.
	
	For $T\in [0, n/2)$, if $b/c<\frac{4n}{\beta_{\text{sig}}(n+2)}$, then the selection gradient $D(x)$ is always negative and there are no singular strategies. If $b/c \in \left(\frac{4n}{\beta_{\text{sig}}(n+2)}, \frac{n(1+e^{\beta_{\text{sig}} T})^2}{\beta_{\text{sig}}(n+2)e^{\beta_{\text{sig}} T}}\right)$, both singular strategies lie in the interval [0, 1]. When $b/c \in \left(\frac{n(1+e^{\beta_{\text{sig}} T})^2}{\beta_{\text{sig}}(n+2)e^{\beta_{\text{sig}} T}}, \frac{n(1+e^{\beta_{\text{sig}}(T-n)})^{2}}{\beta_{\text{sig}}(n+2)e^{\beta_{\text{sig}}(T-n)}}\right)$, only the ESS $x_1^*$ satisfies $x \in [0, 1]$, while $x_2^* < 0$. For $b/c>\frac{n(1+e^{\beta_{\text{sig}}(T-n)})^{2}}{\beta_{\text{sig}}(n+2)e^{\beta_{\text{sig}}(T-n)}}$, $D(x)$ is constantly positive for all $x \in [0, 1]$ and there are no singular strategies.
	
	For $T\in [n/2, n]$, similarly, if $b/c<\frac{4n}{\beta_{\text{sig}}(n+2)}$, then the selection gradient $D(x)$ is always negative and there are no singular strategies. If $b/c\in\left(\frac{4n}{\beta_{\text{sig}}(n+2)}, \frac{n(1+e^{\beta_{\text{sig}}(T-n)})^{2}}{\beta_{\text{sig}}(n+2)e^{\beta_{\text{sig}}(T-n)}}\right)$, both singular strategies lie in the interval [0, 1]. When $b/c\in\left(\frac{n(1+e^{\beta_{\text{sig}}(T-n)})^{2}}{\beta_{\text{sig}}(n+2)e^{\beta_{\text{sig}}(T-n)}}, \\\frac{n(1+e^{\beta_{\text{sig}} T})^2}{\beta_{\text{sig}}(n+2)e^{\beta_{\text{sig}} T}}\right)$, only the repellor $x_2^*$ satisfies $x\in [0, 1]$ and $x_1^* < 0$. And for $b/c>\frac{n(1+e^{\beta_{\text{sig}} T})^2}{\beta_{\text{sig}}(n+2)e^{\beta_{\text{sig}} T}}$, $D(x)$ is constantly positive for all $x\in [0, 1]$ and there are no singular strategies.
	\section*{Acknowledgments}

	This research was supported by the National Natural Science Foundation of China (Grants No. 62473081 and No. 62036002), the Sichuan Science and Technology Program (Grant No. 2024NSFSC0436), and the National Research, Development and Innovation Office under Grant No.~K142948.
	
	\section*{Author Declarations}
	\subsection*{Conflict of Interest}
	The authors have no conflicts to disclose.

\end{document}